\documentclass[11pt,a4paper,table]{article}
\usepackage[bitstream-charter]{mathdesign}

\usepackage{XCharter}
\usepackage{microtype}
\usepackage[T1]{fontenc}
\usepackage[utf8]{inputenc}
\usepackage[british]{babel}
\usepackage{jheppub}
\usepackage{tikz}
\usepackage{slashed}
\usepackage{placeins}
\usepackage[compat=1.1.0]{tikz-feynhand}
\usepackage{comment}
\usepackage{xfrac}
\usepackage{tikz}

\definecolor{MathBlue}{rgb}{0.368417, 0.506779, 0.709798}
\definecolor{MathYellow}{rgb}{0.880722, 0.611041, 0.142051}
\definecolor{MathGreen}{rgb}{0.560181, 0.691569, 0.194885}
\definecolor{MathRed}{rgb}{0.922526, 0.385626, 0.209179}
\definecolor{MathViolet}{rgb}{0.528488, 0.470624, 0.701351}
\definecolor{LightGray}{gray}{0.91}
\definecolor{LightBlue}{rgb}{0.87, 0.94, 1}

\newcommand{\eq}[1]{\eqref{eq:#1}}
\newcommand{\fig}[1]{Fig.~\ref{fig:#1}}
\newcommand{\tab}[1]{Tab.~\ref{tab:#1}}
\newcommand{\Sec}[1]{Sec.~\ref{sec:#1}}

\newcommand{\tr}{\mathrm{tr}\!}

\title{Effective Potential and Vacuum Stability in the Litim-Sannino Model}
\author[a,b]{Tom Steudtner }
\affiliation[a]{TU Dortmund University, Department of Physics, Otto-Hahn-Str.4, D-44227 Dortmund, Germany}
\affiliation[b]{Department of Physics, University of Cincinnati, Cincinnati, OH 45221, USA}
\emailAdd{tom2.steudtner@tu-dortmund.de}
\date{\today}

\abstract{
We revisit the scalar potential in the Litim-Sannino model. 
We compute for the first time the full quantum corrections to the classical potential and show that they significantly ameliorate the stability analysis at the UV fixed point. 
The quantum effective potential is computed at two-loop order and the numerical precision is further improved using resummations and parameter optimisations. 
As a result, we find a consistent widening of the UV conformal window across various approximations.
}

\begin{document}

\maketitle
\section{Introduction}

Historically the paradigm of asymptotic freedom~\cite{Gross:1973id,Politzer:1973fx} has been a centre piece for constructing UV-complete four-dimensional QFTs. 
More recently, a complementary path was discovered 
with the Litim-Sannino model~\cite{Litim:2014uca} and its equivalent formulations~\cite{Bond:2019npq}.
The crucial difference is that these QFTs approach conformality via an interacting UV fixed point in the high energy limit.
And yet, the Litim-Sannino model retains all benefits of asymptotic freedom: both are consistently defined in the deep UV, cutoff-free, and their predictivity is enhanced by the applicability of perturbation theory.
This has enabled a wide range of studies and model building applications, see e.g.~\cite{Litim:2014uca,Sannino:2014lxa,Litim:2015iea,Nielsen:2015una,Rischke:2015mea,Codello:2016muj,Bond:2017wut,Dondi:2017civ,Bond:2017lnq,Bond:2017suy,Abel:2017ujy,Barducci:2018ysr,Bond:2019npq,Hiller:2019mou,Hiller:2019tvg,Hiller:2020fbu,Bissmann:2020lge,Bause:2021prv,Bond:2022xvr,Hiller:2022hgt,Hiller:2022rla}.
In particular, the UV conformal window has been investigated with increasingly high precision~\cite{Bond:2017tbw,Bond:2021tgu,Litim:2023tym,Bednyakov:2023asy}, zoning in on the mechanism causing the eventual demise of the UV fixed point.
These studies have systematically uncovered that the conformal window is entirely within a perturbative regime, and that an instability of the scalar potential is one of the most plausible explanation for its upper end.
Therefore, a dedicated investigation of vacuum stability is vital to explore the conformal window of the Litim-Sannino model. 

While the fixed-point analysis was conducted up to high loop orders, the vacuum stability analysis employed in previous works~\cite{Bond:2017tbw,Bond:2021tgu,Litim:2023tym,Bednyakov:2023asy} is actually a classical one, as first discussed in \cite{Paterson:1980fc}. 
Quantum corrections have been investigated in \cite{Litim:2015iea} by employing a renormalisation group (RG) resummation of the classical potential. This was sufficient to match the leading-order precision of the UV fixed point at the time. Some aspects of symmetry breaking were later studied in \cite{Abel:2017ujy}. 
In this work, we demonstrate that the (semi-) classical approximation of the potential is not on par with the precision of the latest fixed-point results~\cite{Litim:2023tym}. We then extend the scalar stability analysis by computing the effective potential, which fully accounts for quantum corrections.

The quantum effective potential was originally put forward in \cite{Coleman:1973jx} and further developed in \cite{Jackiw:1974cv}. A comprehensive review of early works is found in \cite{Sher:1988mj}.
As of today, generalised formul\ae{} to compute the effective potential for any renormalisable QFT are available up to three loops \cite{Martin:2001vx,Martin:2017lqn} in Landau gauge, as well as at two loop in more general choices of gauge fixing~\cite{Martin:2018emo}.
Since the potential is an observable and overall RG invariant, these fixed-order results may be enhanced by RG resummations, see e.g.~\cite{Kastening:1991gv,Ford:1992mv,Bando:1992np,Bando:1992wy,Andreassen:2014eha,Andreassen:2014gha}.
On the other hand, the potential depends on classical background-field variables, which may introduce an explicit dependence on gauge-fixing parameters~\cite{Nielsen:1975fs, Fukuda:1975di, Aitchison:1983ns}.

This work is structured as follows: \Sec{LiSa} gives a brief introduction to the Litim-Sannino model, our notation, and the stability of its scalar potential at classical level. The computation and discussion of the effective potential follows in \Sec{effective-potential}. \Sec{window} investigates how the enhanced precision influences the conformal window. We close with a final discussion in \Sec{Discussion}.

\section{Litim-Sannino Model}\label{sec:LiSa}

The Litim-Sannino model~\cite{Litim:2014uca} features a non-abelian $SU(N_c)$ gauge group and an unbroken chiral flavour symmetry $U(N_f)_L \times U(N_f)_R$. It contains $N_f$ Dirac fermions $\psi$ in the fundamental representation of the gauge group, as well as $N_f^2$ complex but neutral scalars $\phi$, which are bifundamentals under the global symmetry.
The renormalisable Lagrangian compatible with the symmetry group reads
\begin{equation}\label{eq:lag}
    \begin{aligned}
       \mathcal{L} &=  - \tfrac14\,F^{A\mu\nu} F^A_{\phantom{A}\mu\nu} + \mathcal{L}_\text{gf} + \mathcal{L}_\text{gh} \\
       &\phantom{= }  + \tr\left[\overline{\psi} i \slashed{D}\psi\right] - y\,\tr\left[\overline{\psi} \left( \phi \,\mathcal{P}_R + \phi^\dagger \,\mathcal{P}_L\right)\psi\right]  \\
       &\phantom{= } + \tr\left[\partial^\mu \phi^\dagger \partial_\mu \phi\right]  - m^2\,\tr\left[\phi^\dagger  \phi\right] - u\,\tr\left[\phi^\dagger  \phi\phi^\dagger  \phi\right] - v\,\tr\left[\phi^\dagger  \phi\right] \,\tr\left[\phi^\dagger  \phi\right]\,,
    \end{aligned}
\end{equation}
where traces are taken over flavour and gauge indices while 
$\mathcal{L}_\text{gf}$ and $\mathcal{L}_\text{gh}$ refer to gauge-fixing and ghost terms, respectively.
This family of theories contains gauge interactions, $g$, coupled to the fermions, one real Yukawa coupling, $y$, mediating between the fermionic and scalar sector, as well as two scalar self-interactions $u$ and $v$.

The intriguing aspect of this theory is the occurrence of a weakly interacting UV fixed point. In particular, this fixed point can be brought under strict perturbative control in the planar Veneziano limit~\cite{Veneziano:1976wm}, where $N_{f,c} \to \infty$ while the ratio $N_f/N_c$ is a finite, continuous parameter.
To that end, we introduce the rescaled couplings~\cite{tHooft:1973alw}
\begin{equation}\label{eq:tHooft-couplings}
    \alpha_g= \frac{N_c\,g^2}{(4 \pi)^2}\,, 
    \qquad \alpha_y= \frac{N_c\,y^2}{(4 \pi)^2}\,, 
    \qquad \alpha_u = \frac{N_f\,u}{(4 \pi)^2}\,, 
    \qquad \alpha_v = \frac{N_f^2\,v}{(4 \pi)^2}\,,
\end{equation}   
which are finite in the Veneziano limit and absorb leading powers of $N_{f,c}$. Thus, the explicit dependence on $N_{f,c}$ is traded for a single expansion parameter 
\begin{equation}  
    \epsilon =  \frac{N_f}{N_c} - \frac{11}2\,,
\end{equation}
which is continuously tunable within the range $[-\tfrac{11}2,\infty)$.
The couplings at the UV fixed point can be expanded as power series in $\epsilon$ (conformal expansion) 
\begin{equation}\label{eq:conformal-expansion}
    \alpha_i^* = \sum_{\ell=1}^\infty \alpha_i^{(\ell)} \epsilon^\ell \,,
\end{equation}
Thus, if $\epsilon$ is small but positive, the UV fixed point is perturbative and its existence guaranteed.
The coefficients $\alpha_i^{(\ell)}$ in \eq{conformal-expansion} only receive contributions up to $(\ell+1)$ loops from the gauge- as well as $\ell$ loops from the Yukawa and quartic $\beta$ functions. 
Currently, the coefficients are determined up to $\ell = 3$~\cite{Litim:2023tym}.

The UV fixed point prevails within a parameter range $(0,\,\epsilon_\text{max}]$, known as the UV conformal window. Its extent and the dynamics constraining it have been investigated in~\cite{Bond:2017tbw,Bond:2021tgu,Litim:2023tym,Bednyakov:2023asy}.
The primary reason for the demise of the UV fixed point appears to be vacuum instability, i.e. scalar potential is not bounded from below at the UV fixed point.

The classical stability of the potential has been considered in~\cite{Paterson:1980fc} and will be reviewed below. 
The potential consists only a classical field $\bar{\phi}$, which can be reduced by the applying the $U(N_f)_L \times U(N_f)_R$ global flavour symmetry without loss of generality. In particular, the complex matrix field $\bar{\phi}$ can be diagonalised with real components $h_i$:
\begin{equation}
    \bar{\phi} = \text{diag}(h_1,\,\dots,\, h_{N_f})\,.
\end{equation}
In order to check if the potential is bounded from below, the asymptotic limit of large field values $|h_i| \to \infty$  is of interest. 
Thus, only the quartic part of the classical potential is relevant, which simplifies to
\begin{equation}
    V = u\,\tr\left[\bar{\phi}^\dagger  \bar{\phi}\bar{\phi}^\dagger  \bar{\phi}\right] + v\,\tr\left[\bar{\phi}^\dagger  \bar{\phi}\right] \,\tr\left[\bar{\phi}^\dagger  \bar{\phi}\right] = u\,\sum_i h_i^4 + v\,\sum_{i,j} h_i^2 \,h_j^2 \,.
\end{equation}
For the potential is bounded from below, the quartic part has to be positive for all possible directions $(h_1,\,\dots,\, h_{N_f})^\intercal$ in the space of real moduli.  The initial normalisation of the moduli vector is irrelevant we can fix it by $\sum_i h_i^2  = \mathcal{N}$.
Using this condition, the minimum of the quartic potential is found by solving
\begin{equation}\label{eq:moduli}
   \frac{\partial}{\partial h_k} \left(V - \Lambda \left[\mathcal{N} - \sum_i h_i^2\right]\right) = 4 h_k \left( u\,h_k^2 + v\,\mathcal{N} + \tfrac12\Lambda\right) = 0\,.
\end{equation}
where $\Lambda$ is a Lagrange parameter. The solution of \eq{moduli} implies that each modulus $h_k$ is either $0$ or has a non-vanishing value $h_k = h$, which is universal for all moduli. Assuming that $n$ out of the $N_f$ moduli are non-zero, the quartic potential becomes
\begin{equation}\label{eq:pot-moduli}
    V_\text{cl} = h^4 \left[ n\,u + n^2\,v\right] \,.
\end{equation}
As the potential is manifestly stable (unstable) for $u,v \gtrless 0$, sufficient stability criteria are found for both couplings have opposite signs. The value of $n$ minimising a positive potential \eq{pot-moduli} depends on the sign of each coupling, leading to the stability conditions
\begin{equation}
    \begin{aligned}
     &u > 0 \geq v\,: \quad  & N_f \, u + N_f^2 \, v > 0\,,\\
     &u \leq 0 \leq v\,: \quad   & u + v > 0\,. 
    \end{aligned}
\end{equation}
However, only the first condition is compatible with the Veneziano limit 
\begin{equation}\label{eq:cl-stab}
    \alpha_u > 0\,, \quad \alpha_u + \alpha_v > 0\,,
\end{equation}
while the second one is voided.
Thus, we arrive at the expression
\begin{equation}
    \frac{V_\text{cl}}{(4\pi)^{2}} = \left[\alpha_u + \alpha_v\right] h^4\,
\end{equation}
for the classical potential. In the next section, we review and compute quantum corrections to this potential, extending the results of ~\cite{Litim:2015iea}.

\section{Effective Potential}\label{sec:effective-potential}

\subsection{Fixed-Order Potential}\label{sec:quantum-potential}
Now, we extend the stability consideration to quantum level using the effective potential. For a more detailed introduction see e.g.~\cite{Coleman:1973jx,Jackiw:1974cv,Sher:1988mj}.
Starting from the generating functional of connected Green's functions $W[J]$ 
\begin{equation}
    e^{i W[J]} = \int \mathcal{D}\phi \ e^{i \int \mathrm{d}^4 x \ \left[ \mathcal{L}(\phi) + J \phi \right] }\,,
\end{equation}
the effective action is given by a Legendre transformation
\begin{equation} \label{eq:legendre}
    \Gamma[\bar{\phi}] = W[J] - \int \mathrm{d}^4 x \  J \bar{\phi}
\end{equation}
with a classical field variable $\bar{\phi} = \langle \phi \rangle = \frac{\delta W[J]}{\delta J}$.
Note that the procedure \eq{legendre} is equivalent to expanding each field  $\phi$ around a classical background $\phi \mapsto \phi + \bar\phi$ and integrating out the quantum fluctuations $\phi$
\begin{equation}\label{eq:def-Gamma}
    e^{i \Gamma[\bar{\phi}]} = \int \mathcal{D}\phi \ e^{i \int \mathrm{d}^4 x \ \left[\mathcal{L}(\bar\phi + \phi) + J \phi \right]} \,.
\end{equation}
The parameter $J$ is now a functional of the background field and defined via $J(\bar\phi) = - \frac{\delta \Gamma[\bar\phi]}{\delta \bar{\phi}}$, defined such that $\Gamma[\bar\phi]$ has a minimum at $\bar\phi = 0$. 
The $n$th functional derivative of $\Gamma[\bar\phi]$ with respect to $\bar\phi$ yields the sum of all connected, 1-particle irreducible $n$-point graphs when evaluated at $\bar\phi = 0$.
The effective potential $V_\text{eff}(\bar{\phi})$ is part of $\Gamma[\bar\phi]$ with vanishing momenta
\begin{equation}
    \Gamma[\bar\phi] = \int \mathrm{d}^4 x \left[ - V_\text{eff}(\bar{\phi}) + \mathcal{O}(\partial\bar{\phi})\right]\,,
\end{equation}
and includes the classical potential as well as quantum corrections. For determining $V_\text{eff}$, it is sufficient to choose the background field to be constant. Thus, for computational purposes it is convenient to evaluate $\Gamma[\bar\phi]$ at $\bar{\phi}(x) = \varphi$, 
which amounts to the shift of the classical field by a constant value $\varphi$ in the action. Subsequently, $\varphi$ can be absorbed into the masses and couplings of the theory and hence resummed in loop computations. 

There are two approaches to obtain the potential:
\begin{enumerate}
    \item The effective action, $\Gamma[\varphi]$ may be computed directly from \eq{def-Gamma}~ \cite{Jackiw:1974cv,Fukuda:1974ey}, by integrating out the field $\phi(x)$, which yields
    \begin{equation}\label{eq:Veff-vac}
        V_\text{eff}(\varphi) = V_\text{cl}(\varphi) - \frac{i}2 \int \frac{\mathrm{d}^4 k}{(2\pi)^4} \, \ln \left[\det  i\mathcal{P}_{\phi}(k,\varphi)\right] + \left\langle\exp{\left(i\int d^4 x \ \mathcal{L}_\text{int}\right)}\right\rangle 
    \end{equation}
    for the effective potential.
    Here $V_\text{cl}$ denotes the classical (tree-level) potential, $\mathcal{P}_{\phi}$ the propagator of the quantum field $\phi$, and $\mathcal{L}_\text{int}$ is Lagrangian of cubic and higher interaction terms in $\phi$, all of which containing the background-field variable $\varphi$.
    Note that the contribution $ J \phi$ in \eq{def-Gamma} subtracts all tadpole diagrams.
    The last term of \eq{Veff-vac} consists of all vacuum diagrams at two-loops and higher. See~\cite{Jackiw:1974cv} for more details.
    \item Alternatively~\cite{Lee:1974fj}, one may instead use the relation
    \begin{equation}
        \left.\frac{\delta \Gamma[\bar\phi]}{\delta \bar\phi}\right|_{\bar{\phi} = \varphi}  = - \frac{\partial V_\text{eff}}{\partial \varphi}\,.
    \end{equation}
    Here the left hand side is by definition the 1-particle irreducible tadpole of the quantum field $\phi$ in the original theory where the constant shift $\phi \to \phi + \varphi$ was introduced.
    Thus, $\varphi$ remains as a parameter which has to be integrated in order to obtain the effective potential.
\end{enumerate}

Note that since $\varphi$ is classical, it can be defined in a minimal form, reduced by the application of all anomaly-free global symmetries without loss of generality. In our specific case, this allows us to choose $\varphi = \text{diag}(h_1,\,\dots,\,h_{N_f})$~\cite{Paterson:1980fc} in accordance with the discussion of~\Sec{LiSa}. Since we are only interested in the Veneziano limit, we can further simplify 
\begin{equation}
    \varphi_{ij} = h \,\delta_{ij}\,,
\end{equation}
which corresponds to the ground state corresponding to \eq{cl-stab}.
By absorbing the classical field $h$ into the couplings of quantum fields, we can resum its effect in loop computations. 
This softly breaks the global symmetry group as it introduces additional superrenormalisable couplings, including mass terms. In our particular case, the $U(N_f)_L \times U(N_f)_R$ symmetry reduces to the vectorial subgroup $U(N_f)_V$, while the axial part is broken by the classical field. Overall, the complex scalar decomposes into the classical background field, a real singlet scalar $S$ and pseudo-scalar $\Tilde{S}$, as well as the adjoint fields $R$ and $I$ 
\begin{equation}\label{eq:trafo}
     \phi_{jk} = h\,\delta_{jk}  + \frac{\delta_{jk}}{\sqrt{2 N_f} } \bigg[S + i\, \Tilde{S}\bigg] + R_{jk} + i I_{jk} 
\end{equation}
with the respective mass terms
\begin{equation}\label{eq:masses}
\begin{aligned}
        m_{R}^2 &= m^2 + 2 h^2 \left( 3\,u + N_f \, v\right)\,,\\
        m_{I}^2 &= m^2 + 2 h^2 \left( u + N_f \, v\right)\,,\\
        m_{S}^2 &= m^2 + 6 h^2 \left( u + N_f \, v\right)\,,\\
        m_{\Tilde{S}}^2 &= m^2 + 2 h^2 \left( u + N_f \, v\right)\,,\\
        m_\psi^2 &= y^2\,h^2\,.
\end{aligned}
\end{equation}
Moreover, eq.~\eq{trafo} generates a tadpole interaction
\begin{equation}
    - \mathcal{L}_\text{Tad} = \sqrt{2 N_f} h \left[m^2 + 2 h^2 \left(u + v\,N_f\right) \right]\,S\,.
\end{equation}
The decomposition of the scalar field leads to a plethora of Yukawa
\begin{equation}
    - \mathcal{L}_\text{Yuk} = \frac{y}{\sqrt{2 N_f}}\overline{\psi}_{ia}\left(S + i \gamma_5\,\Tilde{S} \right)\psi_{ia} + y\,\overline{\psi}_{ia}\left(R_{ij} + i \gamma_5\,I_{ij} \right)\psi_{ja}\,,
\end{equation}
scalar cubic
\begin{equation}
    \begin{aligned}
      h^{-1}  V_\text{Cubic} &=  4   u \,\tr\left[RRR + RII\right] \\
            \phantom{= \ } & + 2 \sqrt{\frac{2}{N_f}}  (3\, u + N_f\,v) \, S \, \tr\,R R  +2 \sqrt{\frac{2}{N_f}}   ( u + N_f\,v) \, S \, \tr\,I I \\
       \phantom{= \ } & + 4 \sqrt{\frac{2}{N_f}}   \, u \, \Tilde{S} \, \tr\,R I   + \sqrt{\frac{2}{N_f}}   (u + N_f\,v) \, S\left(S^2 + \Tilde{S}^2\right)    \\
    \end{aligned}
\end{equation}
and quartic interactions
\begin{equation}
    \begin{aligned}
      V_\text{Quartic} &= v \,\tr^2\left[RR + II\right] + u\,\tr \left[RRRR + IIII + 4\,RRII - 2\,RIRI\right] \\
      &\phantom{= \ } + 2\sqrt{\frac{2}{N_f}}\,u\,\left[S\,\tr\,\left(RRR + RII\right) + \tilde{S}\,\tr\,\left(IRR + III\right) \right] \\
      &\phantom{= \ } + \left(v + \frac{3 u}{N_f}\right) \left[S^2 \,\tr\, RR  + \Tilde{S}^2 \,\tr\, II \right] + \left(v + \frac{u}{N_f}\right) \left[S^2 \,\tr\, II  + \Tilde{S}^2 \,\tr\, RR \right] \\
      &\phantom{= \ } + \frac{4 u}{N_f} \, S\Tilde{S} \,\tr\,RI +  \frac{1}{4} \left(v + \frac{u}{N_f}\right) \left(S^2 + \Tilde{S}^2\right)^2\,.
    \end{aligned}
\end{equation}
 
In the Veneziano limit, it is expected that merely the scalars $R$ and $I$ are contributing to loop interactions. 
The same picture emerges for sister theories of the Litim-Sannino model, featuring orthogonal and symplectic gauge groups, Majorana fermions, and a single $SU(N_f)$ flavour symmetry~\cite{Bond:2019npq}. 
In the orthogonal case, the scalar $\phi$ is in the symmetric representation of $SU(N_f)$. Breaking the diagonal component breaks the global symmetry down to $SO(N_f)$, with $R$ and $I$ being in the symmetric representation. In the symplectic case, scalars $\phi$ are in the antisymmetric representation of $SU(N_f)$ while $R$, $I$ are in the antisymmetric representation of $Sp(N_f)$ after the breaking.
An overview is found in \tab{triality}.

\begin{table}[h]
    \centering
    \rowcolors{2}{LightGray}{}
    \begin{tabular}{|l|c|c|c|}
         \hline
         \rowcolor{LightBlue}
            & Dirac $SU(N_c)$ & Majorana $SO(N_c)$ & Majorana $Sp(N_c)$ \\
        \hline
        Unbroken symmetry & $SU(N_f)_L \times SU(N_f)_R $ & $SU(N_f)$ & $SU(N_f)$ \\
        Weyl fermions & $2 N_c N_f$ & $N_c N_f$ & $N_c N_f$ \\
        Scalar DOF $\phi$ & $2 N_f^2$ & $N_f (N_f + 1) $ & $N_f (N_f - 1) $ \\
        \hline 
        Remnant Symmetry & $ SU(N_f)_V$ & $ SO(N_f)$ & $Sp(N_f)$ \\
        Weyl fermions & $2 N_c N_f$ & $N_c N_f$ & $N_c N_f$ \\
        Scalar DOF $R$, $I$ & $N_f^2 -1$ & $\tfrac12 N_f (N_f + 1) - 1$ & $\tfrac12 N_f (N_f - 1) - 1$ \\
        Scalar DOF $S$, $P$ & $1$ & $1$ & $1$  \\
        \hline
    \end{tabular}
    \caption{Overview of fermionic and scalar degrees of freedom with and without symmetry breaking for the Litim-Sannino model and its sister theories~\cite{Bond:2019npq}.}
    \label{tab:triality}
\end{table}

\subsection{Renormalisation Group Improvement}\label{sec:RG-improvement}
The effective potential depends on renormalisation group scale $\mu$ explicitly and implicitly through the running of the background field $h$, couplings $\alpha_i$, mass $m^2$ and the $R_\xi$ gauge fixing parameter $\xi$. However, as the potential is overall renormalisation group invariant and follows a Callan-Symanzik relation
\begin{equation}
    \frac{\mathrm{d}}{\mathrm{d} \ln \mu} V_\text{eff}= \left(\frac{\partial}{\partial \ln \mu} + \gamma_h h\, \frac{\partial}{\partial h} + \beta_j \frac{\partial}{\partial \alpha_j} + \beta_\xi \frac{\partial}{\partial \xi}  + \gamma_{m^2} \frac{\partial}{\partial \ln m^2} \right) V_\text{eff} = 0\,.
\end{equation}
Moreover, we can make the ansatz
\begin{equation}
\frac{V_\text{eff}}{(4\pi)^2} = \alpha_h (z,\kappa,\alpha_i,\xi) \ h^4, \quad \text{ where } \quad z = \frac{(4\pi)^2}{N_f}\frac{h^2}{\mu^2} \quad \text{ and } \quad \kappa = \frac{m^2}{\mu^2}\,,
\end{equation}
which retains higher orders of the classical field in the variable $z$. In the same manner, the scalar mass parameter is swapped for the dimensionless variable $\kappa$ without loss of generality. The potential is bounded from below if $\alpha_h > 0$ for all values of $z$.
The RG invariance condition then reads
\begin{equation}
    \left(- \frac{2 \gamma_h}{1 - \gamma_h} + \frac{\partial}{\partial \ln z} + \frac12 \frac{\beta_j}{1 - \gamma_h} \frac{\partial}{\partial \alpha_j} + \frac12 \frac{\beta_\xi}{1 - \gamma_h} \frac{\partial}{\partial \xi} + \frac{\kappa}2 \frac{\gamma_{m^2} - 2}{1 - \gamma_h} \frac{\partial}{\partial \kappa}\right) \alpha_h = 0\,.
\end{equation}
Note that the term encoding the running of the gauge fixing parameter can always be switched off by choosing the Landau gauge $\xi = 0$, as $\beta_\xi \propto \xi$.
Due to the RG invariance, it is sufficient obtain a solution of the effective potential close to the fixed point in order to argue about vacuum stability.

At the fixed point, $\beta$ functions as well as $\kappa$ vanish, while $z$ remains a tunable parameter, containing the background field dependence.
We obtain
\begin{equation}
    0 = \left(- \frac{2 \gamma_h^*}{1 - \gamma_h^*} + \frac{\partial}{\partial \ln z} \right) \alpha_h (z,0,\alpha_i^*,\xi^*) \,.
\end{equation}
Solving this equation and expressing $\alpha_i^*$, $\xi^*$ through their $\epsilon$ expansion yields a resummed $h$-dependence
\begin{equation}\label{eq:rg-improved_lambda}
    \alpha_h(z,\,0,\,\alpha_i^*,\,\xi^*) =  \alpha_w^*(z_0,\,\epsilon)\, \left(\frac{z}{z_0}\right)^{2 \gamma_h^* /(1 - \gamma_h^*)}\,,
\end{equation}
where $z_0$ is chosen as an arbitrary positive parameter. Obviously, a parameter change in $z_0$ amounts to 
\begin{equation}
    \alpha_w^*(z_0,\,\epsilon) = \left(\frac{z_0}{z_0'}\right)^{2 \gamma_h^* /(1 - \gamma_h^*)} \alpha_w^*(z_0',\,\epsilon)\,.
\end{equation}
Thus, terms in the  effective potential with higher orders in $h$, which are contained in $z$, are resummed in the last factor of eq.~\eq{rg-improved_lambda}.
Most notably, this term is always positive. Hence the boundedness from below of the quantum effective potential at the fixed point is determined by the sign of $\alpha_w^*$ and independent of the field values $z$.
In particular, $\alpha_w^*$ can be computed in perturbation theory by calculating the effective potential in a loop expansion at $z=z_0$:
\begin{equation}\label{eq:alpha_lambda}
  \alpha_w^* =  \sum_{\ell=0}^\infty \alpha_w^{(\ell)} > 0\,.
\end{equation}

Importantly, the stability of the effective potential does not depend on the choice of the parameter $z_0$, which can be fixed in a way that optimises the perturbative expansion.
In particular, logarithmic terms $\propto \epsilon^2 \ln \epsilon/z_0$ are already present at one-loop in $\alpha_w^*$. For e.g. the trivial choice $z_0 = 1$, such terms are under perturbative control as they vanish for $\epsilon \to 0$. However, these logarithms become much more sizeable than non-logarthmic contributions in the limit $\epsilon \to 0$. On the other hand, the choice $z_0 = \epsilon$ eliminates all terms  $\propto \ln \epsilon/z_0$, which improves the perturbative convergence for every order in $\alpha_w^{(\ell)}$. We will return to this opportunity in \Sec{window}.

 The leading term in \eq{alpha_lambda} represents the tree-level of the effective potential
\begin{equation}\label{eq:lambda0}
    \alpha_w^{(0)} = \alpha_u^* + \alpha_v^*\,,
\end{equation}
which coincides with the classical potential and is independent of $z_0$. In \cite{Litim:2015iea,Bond:2017tbw,Bond:2021tgu,Litim:2023tym,Bednyakov:2023asy}, the condition $\alpha_w^{(0)} > 0$ has been adopted as a criterion for vacuum stability. 
This rests on the argument in \cite{Litim:2015iea} that the classical potential $\alpha_w^{(0)}$ controls the sign of leading-log resummations of the background field in the effective potential. 
Ideed, using the loop expansion $\gamma_h^* = \sum_{\ell=1}^\infty \gamma_h^{(\ell)}$, the resummation \eq{rg-improved_lambda} at the fixed point reads  
\begin{equation}\label{eq:leading-log}
    \alpha_h^* = \sum_{n=0}^\infty  \frac{2^n}{n!} \left[\ln \frac{z}{z_0}\right]^n\left\{\ \alpha_w^{(0)}\left(\gamma_h^{(1)}\right)^n 
    + \left[\alpha_w^{(1)} \gamma_h^{(1)} + n \,\alpha_w^{(0)}\left(\gamma_h^{(2)} + \gamma_h^{(1)\,2} \right)\right] \left(\gamma_h^{(1)}\right)^{n-1} + \dots \right\}.
\end{equation}
The first term in the braces of \eq{leading-log} is the leading-log (LL) contribution, which is $\propto \alpha_w^{(0)}$. The second term is next-to-leading-log (NLL) and hence suppressed with an additional loop order. In general, powers $\ln^n z$ involve contributions up N$^n$LL which contain $n$-loop quantum corrections to the classical potential $\alpha_w^{(n)}$.
Boundedness from below is determined in the limit $z\to \infty$ where the term $n\to\infty$ in \eq{leading-log} dominates, which means that all loop corrections to $\alpha_w^*$ contribute to its coefficient and thus vacuum stability.
The classical stability condition $\alpha_w^{(0)} > 0$ ensures the positivity of LL term which is na\"ively the largest contribution. 
Although pertubatively suppressed, the subleading-log corrections may still overpower the LL numerically, in particular at the edge of classical stability where $\alpha_w^{(0)} \approx 0$. This scenario is accounted for when considering the full $\alpha_w^*$.
Thus, the focus of this work is the computation of quantum corrections  $\alpha_w^{(\ell>0)}$, in order to improve the prediction on vacuum stability.

 \subsection{Fixed-Point Potential}

In this section, the stability of the scalar potential in the high-energy limit is determined by computing the quantity $\alpha_w$, defined in \eq{rg-improved_lambda} at the UV fixed point.
Since all fixed-point values $\alpha_{g,y,u,v}^*$  of classically marginal couplings are known up to $\propto \epsilon^3$ in the conformal expansion, the effective potential needs to be calculated up to two-loop order to reach the same precision for $\alpha_w$.
The effective potential has been determined by a direct loop computation using the computational setup of \texttt{MaRTIn}~\cite{Brod:2024zaz}.
Both methods outlined in \Sec{quantum-potential} have been pursued and yield identical results. Finally, literature expressions applicable to generic renormalisable QFTs~\cite{Martin:2001vx,Martin:2018emo} have been utilised to cross check the results.
At this loop order, there is no dependence on the gauge fixing.
Concretely, we obtain
 \begin{align}
    &\alpha_w^{(0)} = \alpha_u + \alpha_v\,,\label{eq:al1}\\[.5em]
    &\begin{aligned}
        \alpha_w^{(1)} &= (\alpha_u + \alpha_v)^2 \left[L_I - \tfrac32\right] + (3\,\alpha_u + \alpha_v)^2 \left[L_R - \tfrac32\right] - \left(\tfrac{11}{2} + \epsilon\right) \alpha_y^2 \left[L_F - \tfrac{3}{2} \right]\,,
    \end{aligned}\label{eq:al2}\\[.5em]
    &\begin{aligned}
        \alpha_w^{(2)} &= \left(\tfrac{11}2+  \epsilon\right)^2 \left[ 4 \,\Phi_{RF} -\left(3 - L_F\right)^2 \right] \alpha_y^3  + \left(\tfrac{11}2+  \epsilon\right) \left[ 9 - 8\,L_F + 3\,L_F^2\right] \alpha_y^2\, \alpha_g \\
        &\phantom{=\ } - (11+2\epsilon) \left[
        9  + 4\,L_F  - L_I  - 15 L_R + L_F\left( L_I + 9\,L_R - 5\,L_F\right)\right] \alpha_y^2 \,\alpha_u\\
        &\phantom{=\ } - (11+2\epsilon) \left[
        2  + 2\,L_F  - L_I  - 5 L_R + L_F\left( L_I + 3\,L_R - 2\,L_F\right)\right] \alpha_y^2 \,\alpha_v\\
        &\phantom{=\ } - (11+2\epsilon) \left[(\alpha_u + \alpha_v) \, \Phi_{IF} + 2(3\alpha_u + \alpha_v) \, \Phi_{RF} \right]\alpha_y^2 \\
        &\phantom{=\ } + \left[50 - 10\,L_F^2 + 2(L_F - 2)(9\,L_R + L_I) + \Phi_{IF} + 9\,\Phi_{RF} \right] \alpha_y\,\alpha_u^2\\
        &\phantom{=\ } + \left[10 - 2\,L_F^2 + 2(L_F - 2)(L_R + L_I) + \Phi_{IF} +\Phi_{RF} \right] \alpha_y\,\alpha_v^2\\
        &\phantom{=\ } + \left[40 - 8\,L_F^2 + 4(L_F - 2)(3\,L_R + L_I) + 2\,\Phi_{IF} + 6\,\Phi_{RF} \right] \alpha_y\,\alpha_u\,\alpha_v\\
        &\phantom{=\ } + \left[192 - 243\,S_2 - \Phi_{RI} - 24\,L_I - 168\, L_R + L_I^2 + 18\,L_I L_R + 45\,L_R^2  \right] \alpha_u^3\\
        &\phantom{=\ } + \left[108 - 81\,S_2 - 3\,\Phi_{RI} - 40\,L_I - 104\, L_R + 6\,L_I^2 + 24\,L_I L_R + 30\,L_R^2  \right] \alpha_u^2\,\alpha_v\\
        &\phantom{=\ } + 4(2 - L_I - L_R) (3 - L_I - 2\,L_R ) \alpha_u\,\alpha_v^2 + (2 - L_I - L_R)^2 \alpha_v^3\,.\label{eq:al3}
    \end{aligned}
 \end{align}
For convenience we define the couplings
 \begin{equation}\label{eq:mass-couplings} 
    \alpha_F = \left(\tfrac{11}{2} + \epsilon\right)\alpha_y \,,\quad \alpha_I = 2 (\alpha_u + \alpha_v)\,,\quad \alpha_R = 2 (3\,\alpha_u + \alpha_v)\,,
\end{equation}
which stem from the squared masses \eq{masses} of fermions and adjoint scalars $R$ and $I$ as the unbroken scalar mass parameter $m^2$ is a relevant operator and vanishes at the fixed point $m^2 \to 0$.
We have employed these couplings as arguments for the (poly-) logarithmic terms
 \begin{equation}\label{eq:logs}
    L_X = \ln\left[\alpha_X/z_0\right]\,,\qquad \Phi_{XY} = \Phi\left(\frac{\alpha_X}{4\, \alpha_Y}\right) \,,
 \end{equation}
 where $X,\,Y = F,\,R,\,I$. Note that these contributions are the only dependencies of the parameter $z_0$.
 The $\Phi_{XY}$ are loop functions defined in~\cite{Davydychev:1992mt} in terms of hypergeometric functions
\begin{equation}
    \Phi(z) = 4\,z\left[(2 - \ln 4z)\ {}_2F_1\left(\left.{}^{1,\,1}_{3/2}\right| z \right)
    - \frac{\partial}{\partial a}\ {}_2F_1\left(\left.{}^{a,\,1}_{3/2}\right| z \right) \bigg|_{a=1}
    - \frac{\partial}{\partial c}\ {}_2F_1\left(\left.{}^{1,\,1}_{\ c}\right| z \right) \bigg|_{c=3/2}
    \right]
\end{equation}
or equivalently
\begin{equation}
    \Phi(z) = \left\{\begin{array}{cc}
        \frac{4}{\sqrt{1/z - 1}} \mathrm{Cl}_2 \left(2 \,\mathrm{arcsin} \sqrt{z}\right) & 0 \leq z < 1, \\
        \quad 8 \ln 2 & z = 1, \\
        \frac1{\sqrt{1-1/z}} \left[\frac{\pi^2}{3} - \ln^2 4z + 2 \ln^2 \frac{1- \sqrt{1-1/z}}2 - 4 \mathrm{Li}_2 \left(\frac{1- \sqrt{1-1/z}}2 \right) \right] & z > 1
    \end{array}\right.
\end{equation}
in the non-negative parameter range, where $\mathrm{Li}_2$ is a dilogarithm and $\mathrm{Cl}_2$ denotes the Clausen function.
For large arguments, the function diverges as 
\begin{equation}
    \lim_{z\to \infty} \Phi(z) = \frac{\pi^2}{3} + \ln^2 4z\,.
\end{equation}
 Finally, $S_2$ is a constant with the value 
 \begin{equation}
    S_2 = \tfrac19\Phi\left(\tfrac14\right) = \tfrac{4}{9\sqrt{3}} \mathrm{Cl}_2\left(\tfrac\pi3\right)  \approx 0.260434\dots \,.
 \end{equation}

 In the region where the potential is classically stable $\alpha_u>0$, $\alpha_u + \alpha_v > 0$, the squared masses \eq{masses} of fermions and all scalars are positive.
This changes at the point of classical flatness $\alpha_u + \alpha_v = 0$ where $L_I$ becomes singular.
However, the effective potential remains well-defined
\begin{align}\label{eq:classically-flat}
    &\alpha_w^{(0)} \big|_{\alpha_u + \alpha_v = 0} = 0\,,\\[.5em]
    &\begin{aligned}
        \alpha_w^{(1)}\big|_{\alpha_u + \alpha_v = 0} &= 4\,\alpha_u^2 \left[L_R - \tfrac32\right] - \left(\tfrac{11}{2} + \epsilon\right) \alpha_y^2 \left[L_F - \tfrac{3}{2} \right]\,,
    \end{aligned}\\[.5em]
    &\begin{aligned}
        \alpha_w^{(2)}\big|_{\alpha_u + \alpha_v = 0} &= \left(\tfrac{11}2+  \epsilon\right)^2 \left[ 4 \,\Phi_{RF} -\left(3 - L_F\right)^2 \right] \alpha_y^3  + \left(\tfrac{11}2+  \epsilon\right) \left[ 9 - 8\,L_F + 3\,L_F^2\right] \alpha_y^2\, \alpha_g \\
        &\phantom{=\ } - (11+2\epsilon) \left[
        7  - 10 L_R + L_F\left( 2 + 6\,L_R - 3\,L_F\right) + 4 \Phi_{RF}\right] \alpha_y^2 \,\alpha_u\\
        &\phantom{=\ } + 4 \left[5 + \Phi_{RF} - L_F^2 + 2\, L_R (L_F - 2)\right] \alpha_y\,\alpha_u^2 \\
        &\phantom{=\ } + \left[104 + \tfrac23 \pi^2 - 162 S_2 - 88\,L_R + 24 \,L_R^2\right]\alpha_u^3 \,.
    \end{aligned}
\end{align}

In general, there are two symptoms of the effective potential coupling $\alpha_w$ that indicate the breakdown of the UV fixed point: 
\begin{itemize}
    \item[(i)] $\mathrm{Re}\, \alpha_w < 0$ signals that the potential is not bounded from below and therefore no ground state of finite energy exists.
    \item[(ii)] $\mathrm{Im}\, \alpha_w \neq 0$ is an indication that the effective potential ceases to be convex~\cite{Fujimoto:1982tc}, which suggests an additional minimum with $h \neq 0$. The imaginary part can be interpreted as the decay width between degenerate vacua~\cite{Weinberg:1987vp,Sher:1988mj}.
\end{itemize}
In the classical approximation $\alpha_w = \alpha_w^{(0)}$, both phenomena coincide at $\alpha_u^* + \alpha_v^* < 0$. For one, this suggests that the classical potential is not bounded from below as $h \to \pm \infty$, and there is no ground state of finite energy.
Furthermore, the condition also turns the tree-level mass of the $I$ field tachyonic,  $m_I^2 < 0$, implying  symmetry breaking into two non-trivial vacua and therefore a non-convex potential. 
At loop level, $m_I^2 < 0$ generates an imaginary part through $L_I$, $\Phi_{IF}$ and $\Phi_{RI}$.

However, both mechanisms (i) and (ii) receive corrections at loop level, such that it is not clear \textit{a priori} which one is the dominant effect to herald scalar instability.
In particular, even the imaginary part implied by $m_I^2 < 0$ may potentially be cancelled by resumming higher-order corrections. The question whether $\mathrm{Im}\, \alpha_w \neq 0$ is of physical consequence or merely signals a deficiency of the perturbative expansion is indeed non-trivial.
The parameter region close to vacuum instability is numerically problematic as 
$|m_I^2|$ is very tiny or vanishing, implying large logarithmic contributions $\propto \ln m_I^2$ to the effective potential at each loop order and thus weakening the perturbative convergence. This is known as the Goldstone catastrophe~\cite{Martin:2014bca,Elias-Miro:2014pca,Andreassen:2014eha,Kumar:2016ltb,Braathen:2016cqe}. The impact of the Goldstone catastrophe
can be mitigated by resumming diagrams that correspond to soft contributions to the light or vanishing scalar mass.
This yields corrections to their masses  $\Tilde{m}_I^2 = m_I^2 + \delta m_I^2$. The corrected mass $\Tilde{m}_I^2$ is then used in place of $m_I^2$, especially in logarithms where the numerical impact of small $|m_I^2|$ is large. This corresponds to resumming higher-order terms in the effective potential.
In our case the strategy is implemented by means of the softened coupling
\begin{equation}\label{eq:soft-aI}
    \Tilde{\alpha}_I = \alpha_I + \Delta_I 
\end{equation}
where the shift admits a loop expansion $\Delta_I = \sum_{\ell=1}^\infty \Delta_I^{(\ell)}$.
In order to retain the non-logarithmic terms of the effective potential up to $\propto \epsilon^3$ while also addressing the Goldstone catastrophe with the same precision, we make the replacement $\alpha_I \mapsto \Tilde{\alpha}_I$ in the logarithmic terms, yielding the alternative definitions
\begin{equation}
    L_I = \ln\left[\frac{\alpha_I + \Delta_I}{z_0}\right], \quad 
    \Phi_{IF} = \Phi\left[\frac{\alpha_I + \Delta_I }{ 4 \alpha_F}\right]\,, \quad
    \Phi_{RI} = \Phi\left[\frac{\alpha_R}{4(\alpha_I + \Delta_I)}\right]
\end{equation}
in \eq{al1}--\eq{al3}
as well as the shift
\begin{equation}
    \alpha_w^{(2)} \mapsto \alpha_w^{(2)} - \tfrac12 (\alpha_u + \alpha_v) \Delta_I^{(1)}\,.
\end{equation}
to avoid double counting of resummed contributions.
This mechanism allows to trade the imaginary part of $\alpha_w$ for a real contribution while improving the numerical reliability.
The choice of $\Delta_I$ is not unique but the impact of different prescriptions diminishes with higher loop orders of the effective potential.
The literature~\cite{Martin:2014bca,Elias-Miro:2014pca,Andreassen:2014eha,Kumar:2016ltb,Braathen:2016cqe}, advocates choices related to the self-energy of the Goldstone boson at zero momentum $\Pi_I(0)$, which can be interpreted as a resummation of soft leg corrections to its mass. Self-interactions of the Goldstone bosons are often neglected.
In our case, we choose to resum Yukawa and gauge interactions to implement a sufficiently effective resummation
\begin{equation}
    \Delta_I = \sum_{\ell=1}^\infty \Delta_I^{(\ell)} = \frac{N_f}{h^2} \Pi_I(0) \Big|_{\alpha_{I,R} = 0}\,.
\end{equation}
We have computed the self-energy to two loop order to extract $\Tilde{\alpha}_I$ up to order $\propto \epsilon^3$, and obtain in the loop expansion
\begin{equation}\label{eq:our-DeltaI}
\begin{aligned}
    \Delta^{(1)}_I &= (11 + 2 \epsilon) \alpha_y^2 \left[1 - L_F\right] \,,\\[.5em]
    \Delta^{(2)}_I &= (11 + 2 \epsilon) \alpha_g \alpha_y^2 \left[5 - 5\,L_F + 3\,L_F^ 2\right] - \tfrac12(11 + 2 \epsilon)^2 \alpha_y^3 \left[6 - 5\,L_F + L_F^2 - 4\,\Phi_{RF} \right]\,,
\end{aligned} 
\end{equation}
which corresponds to the diagrams
\newline
\begin{tabular}{ccccc}
    \begin{tikzpicture}[scale=.75,baseline=0em]
        \begin{feynhand}
            \vertex [particle] (e1) at (0em,0em) {}; 
            \vertex [particle] (e2) at (8em,0em) {};
            \vertex [dot] (i1) at (2em,0em) {}; 
            \vertex [dot] (i2) at (6em,0em) {};

            \propag [scalar] (e1) to (i1);
            \propag [scalar] (e2) to (i2);

            \draw (4em,0em) circle [radius=2em];

        \end{feynhand} 
    \end{tikzpicture}&
    \begin{tikzpicture}[scale=.75,baseline=0em]
        \begin{feynhand}
            \vertex [particle] (e1) at (0em,0em) {}; 
            \vertex [particle] (e2) at (8em,0em) {};
            \vertex [dot] (i1) at (2em,0em) {}; 
            \vertex [dot] (i2) at (6em,0em) {}; 
            \vertex [dot] (u3) at (4em,+2em) {}; 
            \vertex [dot] (d3) at (4em,-2em) {}; 

            \propag [scalar] (e1) to (i1);
            \propag [scalar] (e2) to (i2);

            \draw (4em,0em) circle [radius=2em];
            \propag [scalar] (d3) to (u3);

        \end{feynhand}
    \end{tikzpicture} &
    \begin{tikzpicture}[scale=.75,baseline=0em]
        \begin{feynhand}
            \vertex [particle] (e1) at (0em,0em) {}; 
            \vertex [particle] (e2) at (8em,0em) {};
            \vertex [dot] (i1) at (2em,0em) {}; 
            \vertex [dot] (i2) at (6em,0em) {}; 
            \vertex [dot] (ul) at (2.5em, 1.3em) {};
            \vertex [dot] (ur) at (5.5em, 1.3em) {};

            \propag [scalar] (e1) to (i1);
            \propag [scalar] (e2) to (i2);

            \propag [scalar] (ul) to [quarter right] (ur);

            \draw (4em,0em) circle [radius=2em];

        \end{feynhand}
    \end{tikzpicture} &
    \begin{tikzpicture}[scale=.75,baseline=0em]
        \begin{feynhand}
            \vertex [particle] (e1) at (0em,0em) {}; 
            \vertex [particle] (e2) at (8em,0em) {};
            \vertex [dot] (i1) at (2em,0em) {}; 
            \vertex [dot] (i2) at (6em,0em) {}; 
            \vertex [dot] (u3) at (4em,+2em) {}; 
            \vertex [dot] (d3) at (4em,-2em) {}; 

            \propag [scalar] (e1) to (i1);
            \propag [scalar] (e2) to (i2);

            \draw (4em,0em) circle [radius=2em];
            \propag [gluon] (d3) to (u3);

        \end{feynhand}
    \end{tikzpicture}&
    \begin{tikzpicture}[scale=.75,baseline=0em]
        \begin{feynhand}
            \vertex [particle] (e1) at (0em,0em) {}; 
            \vertex [particle] (e2) at (8em,0em) {};
            \vertex [dot] (i1) at (2em,0em) {}; 
            \vertex [dot] (i2) at (6em,0em) {}; 
            \vertex [dot] (ul) at (2.5em, 1.3em) {};
            \vertex [dot] (ur) at (5.5em, 1.3em) {};
            
            \propag [scalar] (e1) to (i1);
            \propag [scalar] (e2) to (i2);

            \propag [gluon] (ul) to [quarter right] (ur);

            \draw (4em,0em) circle [radius=2em];

        \end{feynhand}
    \end{tikzpicture} 

    \end{tabular}.

\subsection{Away from the UV Fixed Point}

As the effective potential is RG invariant, vacuum stability at the UV fixed point also guarantees the potential to be bounded from below in the IR. 
Thus, scalar stability reigns along the weakly coupled trajectory connecting the interacting UV fixed point with IR freedom.
More explicitly, we follow the relevant trajectory
\begin{equation}
   \alpha_g = \alpha_g^*(\epsilon) + \delta,\qquad \alpha_{y,u,v} = \alpha_{y,u,v}^*(\epsilon) + \sum_{n=1}^\infty c_{y,u,v}^{(n)}(\epsilon) \,\delta^n
\end{equation}
which emanates from the fixed point at $\delta = 0$ and leads into the weakly $(\delta < 0)$ or strongly $(\delta > 0)$ coupled regime. The RG flow of the effective coupling 
$\delta$ at leading order is related to the relevant critical exponent $\beta_\delta = \mathrm{d} \delta / \mathrm{d} \ln \mu = \vartheta_1\,\delta + \mathcal{O}(\delta^2)$.
The RG invariance of the effective potential away from the fixed point is encoded in the condition
\begin{equation}
   0 = \left(- \frac{2 \gamma_h}{1 - \gamma_h} + \frac{\partial}{\partial \ln z} + \frac12 \frac{\beta_\delta}{1 - \gamma_h} \frac{\partial}{\partial \delta}  \right) \alpha_h(z,\delta,\epsilon) \,.
\end{equation}
In accordance with \eq{rg-improved_lambda} we apply the ansatz 
\begin{equation}\label{eq:rg-improved_lambda-2}
    \alpha_h(z,\,\delta,\,\epsilon) =  \alpha_w(\delta,\,z_0,\,\epsilon)\, \left(\frac{z}{z_0}\right)^{2 \gamma_h^* /(1 - \gamma_h^*)}\,,
 \end{equation}
 yields a condition 
\begin{equation}
   0 = \left(\frac{\gamma_h - \gamma_h^*}{1 - \gamma_h^*}   - \frac{\beta_\delta}{4} \frac{\partial}{\partial \delta}  \right) \alpha_w(\delta,\,z_0,\,\epsilon)
\end{equation}
that is valid away from the UV fixed point and solved by 
\begin{equation}\label{eq:separatrix-potential}
    \alpha_w  = \alpha_w^* \, \exp\left\{\frac4{1 - \gamma_h^*} \int_0^{\delta} \mathrm{d} \hat{\delta} \, \frac{\gamma_h(\hat{\delta}) - \gamma_h^*}{ \beta_{\hat{\delta}}}\right\}\,.
 \end{equation}
 Eqs.~\eq{rg-improved_lambda-2} and \eq{separatrix-potential} imply that if the effective potential is stable at the UV fixed point through $\alpha_w^* > 0$, it also remains stable away from it.

The scalar mass operator $m^2$ as in \eq{lag} is relevant and vanishes at the UV fixed point. If it is switched on at some IR scale $m^2(\mu_\text{IR}) > 0$,
all scalars eventually decouple and the theory consists of free fermions and vector bosons in the low-energy limit, with an intact $U(N_f)_L \times U(N_f)_R$ global symmetry. If $m^2(\mu_\text{IR}) = 0$, all fields including scalars remain massless until IR freedom is reached.
For $m^2(\mu_\text{IR}) < 0$, spontaneous symmetry breaking occurs and the scalar sector develops a vacuum expectation value $h$ as in~\eq{trafo}.
Other mechanisms to violate the global symmetry in the IR without spoiling the UV fixed point require an explicit breaking by relevant operators~\cite{Abel:2017ujy}.
A dynamical symmetry breaking by turning relevant couplings to be irrelevant via the RG cannot occur in the weak coupling regime.

The other trajectory emanating from the UV fixed point leads away from the weak regime towards strong couplings. Again, spontaneous symmetry breaking occurs if  $m^2(\mu_\text{IR}) < 0$.  
For $m^2(\mu_\text{IR}) = 0$, the theory may reach a putative interacting IR fixed point, a scenario compatible with the observation of fixed-point mergers~\cite{Bond:2017tbw,Bond:2021tgu,Litim:2023tym}.
Alternatively, the IR trajectory may also lead into a strongly coupled regime, where 
an additional scale parameter is generated via dimensional transmutation, and non-perturbative phenomena arise such as the breaking of global symmetries and confinement.
For larger value of $\epsilon$, an indication for non-perturbative effects causing a symmetry breaking in the effective potential is $\gamma_h^* = 1$, which causes a breakdown of the solution \eq{rg-improved_lambda}.

\section{Impact on Conformal Window}\label{sec:window}

In this section we discuss the UV conformal window of the Litim-Sannino model, namely the value $\epsilon_\text{max}$ at which the fixed point ceases to exist. 
Several mechanisms are known that may cause such a disappearance~\cite{Kaplan:2009kr}. As established by previous works~\cite{Bond:2017tbw,Litim:2023tym,Bednyakov:2023asy}, the conformal window appears to be within a weakly-interacting regime $|\alpha_{g,y,u,v}^*(\epsilon)| \ll 1$ and not limited due to strong-coupling effects. Rather, the UV fixed point may disappear into the complex plane due to a fixed-point merger. This occurs when an IR fixed-point solution  exists at $\epsilon <  \epsilon_\text{mer}$ that approaches the UV one with increasing $\epsilon$, causing both solutions to become complex at $\epsilon >  \epsilon_\text{mer}$. 
We distinguish two subcategories of this phenomenon: single- and double-trace mergers. A single-trace merger is characterised by the relevant critical exponent turning marginal
\begin{equation}
  \vartheta_1(\epsilon_\text{mer,1}) = 0\,,  
\end{equation} 
signalling the disappearance of the UV fixed point. Typically, the single-trace couplings $\alpha_{g,y,u}^*(\epsilon < \epsilon_\text{mer,1})$ of the colliding UV and IR fixed point solutions are different before this merger. 
The second type of merger is due to the algebraic decoupling in the planar Veneziano limit. The single-trace RG evolution via $\beta_{g,y,u}$ is independent of the double-trace quartic $\alpha_v$, while its own $\beta$ function is only quadratic in it
\begin{equation}
    \beta_v = f_0 + f_1\, \alpha_v + f_2 \,\alpha_v^2
\end{equation}
up to all loop orders~\cite{Pomoni:2009joh}. Thus, each fixed-point solution $\alpha_{g,y,u}^*$ implies up to two real solutions $\alpha_v^{*\pm}$ for the double-trace quartic with the critical exponents 
\begin{equation}
    \vartheta_3^\pm = \pm \sqrt{f_1^{*2} - 4 f_0^* f_2^*} \,.
\end{equation}
For the UV fixed point, only $\alpha_v^{*+}$ is physical while $\alpha_v^{*-}$ exhibits an unstable potential~\cite{Litim:2023tym}. A double-trace merger occurs when both solutions collide such that 
\begin{equation}
    \alpha_v^{*+} (\epsilon_\text{mer,2}) = \alpha_v^{*-} (\epsilon_\text{mer,2}) \qquad \text{ and } \qquad \vartheta_3^\pm(\epsilon_\text{mer,2}) = 0.
\end{equation}
While there is evidence of a single-trace merger, a double-trace merger seems to be absent in the $\epsilon$-expansion~\cite{Litim:2023tym}. 
However, constraints to the conformal window due to instability of the fixed-point potential, $\epsilon_\text{vac}$, appear to be slightly tighter. To settle which mechanism is more dominant for the upper end of the conformal window, more precision in their determination is required.
Thus, in the following analysis we improve upon previous works~\cite{Litim:2023tym,Bednyakov:2023asy} by bringing the stability analysis on par with available loop orders of the $\beta$ functions.
To do so, we first take stock of all uncertainties for the conformal-window estimate.

\subsection{Uncertainties}
Computing the effective potential at the UV fixed point within a finite order in perturbation theory, we are confronted with the following sources of uncertainty.
\begin{enumerate}
    \item Uncertainty of the fixed-point couplings $\alpha_{g,y,u,v}^*(\epsilon)$ and critical exponents $\vartheta_{1,2,3,4}(\epsilon)$. The state-of-the.art is to extract these quantities from four-loop gauge, three-loop Yukawa and three-loop quartic $\beta$ functions, \texttt{433} in short.
    Thus, $\alpha_{g,y,u,v}^*$ are determined up to the third coefficient $\propto \epsilon^3$ in the conformal expansion and do not receive contributions from higher loops. We will also use various techniques to resum higher powers in $\epsilon$, though the reliability of such approaches will remain an open question and will be used to estimate the residual uncertainty. 
    
    \item Uncertainty from missing loop orders of the effective potential. In order to match the $\propto \epsilon^3$ expansion of the fixed-point couplings, the effective potential is required to two-loop order. Given that the fixed point remains in a perturbative regime, higher-order contributions to the potential should be increasingly irrelevant. To verify this, we compare the impact of various loop orders in this section.
    \item Ambiguity of the Goldstone resummation $\Delta_I$. As discussed earlier, a softened parameter $\tilde{\alpha}_I$~\eq{soft-aI} is introduced to improve the numerical stability through resummation. This procedure moves the hints of instability from the imaginary to the real part of the effective potential. However, $\Delta_I$ is not uniquely defined. Here we keep track on how the choice~\eq{our-DeltaI} alters the real part of the unresummed potential and ensure that $\tilde{\alpha}_I(\epsilon_\text{vac}) > 0$ as $\alpha_w(\epsilon_\text{vac}) = 0$.
    \item Suitable choice of $z_0$. The overall stability estimate is independent of the choice of  $z_0$. This parameter is introduced in \eq{rg-improved_lambda} as a mediator between the exact field resummation and the perturbatively determined coefficient $\alpha_w$ in the effective potential. 
    The convergence of the loop expansion of $\alpha_w$ can be optimised by a smart choice of $z_0$. Moreover, an uncertainty estimate may be obtained by varying $z_0$.     
\end{enumerate}

Let us now develop a strategy regarding the last point.
The parameter $z_0$ only appears within the logarithms \eq{logs}. Thus, a natural choice of $z_0$ would be one of the couplings $\alpha_F$, $\alpha_R$ or $\tilde{\alpha}_I$ to cancel corresponding terms logarithmic terms, or an intermediate value. Values of $z_0$ outside the range spanned by $\alpha_F$, $\alpha_R$ and $\tilde{\alpha}_I$  cause unnecessarily large logarithmic contributions.
At the fixed point the hierarchy
\begin{equation}\label{eq:hierarchy}
    0 < \Tilde{\alpha}_I \ll \alpha_R \approx \epsilon < \alpha_F \ll 1
\end{equation}
usually holds. This is transparently seen in the leading-order expansion
\begin{equation}
    0 < 0.13\,\epsilon \ll 0.92\,\epsilon \approx \epsilon < 1.16\,\epsilon \ll 1\,.
\end{equation}
This suggests that $z_0$ should be chosen within $\alpha_F \geq z_0 \geq \Tilde{\alpha}_I$ to optimize $\alpha_w$.
In this range the largest impact of varying $z_0$ is due to the rift $\tilde{\alpha}_I \ll \alpha_{R,F}$.
Fortunately, logarithmic terms in the effective potential are softened by the couplings in front.

This can be seen for instance in \eq{al2}, which contains both the terms $\propto \alpha_R^2 \ln \alpha_R/z_0$ and $\propto \alpha_I^2 \ln \tilde{\alpha}_I/z_0$. The reader is reminded that by definition \eq{soft-aI}, $\tilde{\alpha}_I$ and $\alpha_I$ differ only by a higher loop expression.
The choice $z_0 \approx \tilde{\alpha}_I$ cancels the logarithm $\propto \alpha_I^2 \ln \tilde{\alpha}_I/z_0$ minimizes. However, as $\alpha_I^2 \ll \alpha_R^2$, the term was already smaller than the other $\alpha_R^2 \ln \alpha_R/z_0$. In fact the latter term $\propto \alpha_R^2 \ln \alpha_R/\tilde{\alpha}_I$ becomes even more sizeable due to the logarithm. Instead, the converse choice $z_0 \approx \alpha_R$ is more meaningful: the logarithm $\propto \alpha_R^2 \ln \alpha_R/z_0$ is traded in for a contribution $\propto \alpha_I^2 \ln \tilde{\alpha}_I/\alpha_R$. While the logarithm is equally large in both cases, the perturbative suppression by the coupling in front is much stronger in the second approach.
Conclusively, $z_0$ should be chosen away from $\tilde{\alpha}_I$ to optimise the perturbativity of loop corrections to the effective potential. 
More precisely, the parameter should be varied in the range of remaining logarithmic scales
\begin{equation}\label{eq:z0-uncertainty}
    \alpha_F \geq z_0 \geq \alpha_R \,,
\end{equation}
which also yields an uncertainty estimate. 
For the perturbative expansion of the effective potential to be reliable, we expect that for each loop order the uncertainty range due to \eq{z0-uncertainty} is decreasing and higher-loop ranges are roughly nested within lower-order ones.

\subsection{Conformal Expansion}

In this section, we investigate the implications of scalar stability from the $\epsilon$ expansion of the effective potential.
The expansion is reliable up to $\propto \epsilon^3$, after which higher loop orders in the $\beta$ functions and the effective potential are required.
We start by recalling the classical stability condition $ \alpha_w^{(0)} > 0$ with
\begin{equation}
    \alpha_w^{(0)} = \alpha_u^* + \alpha_v^* \approx  0.0625\,\epsilon - 0.1915\,\epsilon^2 - 1.6200\,\epsilon^3 + \mathcal{O}(\epsilon^4)\,.\\
\end{equation}
This result is improved by taking quantum corrections into account
\begin{equation}\label{eq:a_lambda-eps}
    \begin{aligned}
    \alpha_w   \approx   &\ 0.0625\,\epsilon \\
    &- \left(0.2127 + 0.0263\,\ln \frac{\epsilon}{z_0}\right)\epsilon^2 \\
    &- \left(1.0536 - 0.0372 \,\ln \frac{\epsilon}{z_0} - 0.0055\,\ln^2 \frac{\epsilon}{z_0}\right)\epsilon^3 + \mathcal{O}(\epsilon^4)\,.
    \end{aligned}
\end{equation}
At order $\propto \epsilon$, both conditions are equivalent and the contribution is manifestly positive, pointing towards stability. At higher orders the quantum effective potential depends on $z_0$.
For $z_0 = \epsilon$, the $\propto \epsilon^2$ correction to the effective potential is negative, indicating an instability at a certain value for $\epsilon_\text{vac}$. Moreover, the coefficient is more negative than the classical one, implying that the conformal window may be smaller once quantum corrections are included.
The $\propto \epsilon^3$ coefficients are again negative in both cases, reinforcing the evidence for the loss of vacuum stability. However, the quantum effective coefficient is larger than in the classical potential, thus relaxing the constraint on the conformal window.

\begin{figure}[h]
    \centering
    \includegraphics[width=0.66\textwidth]{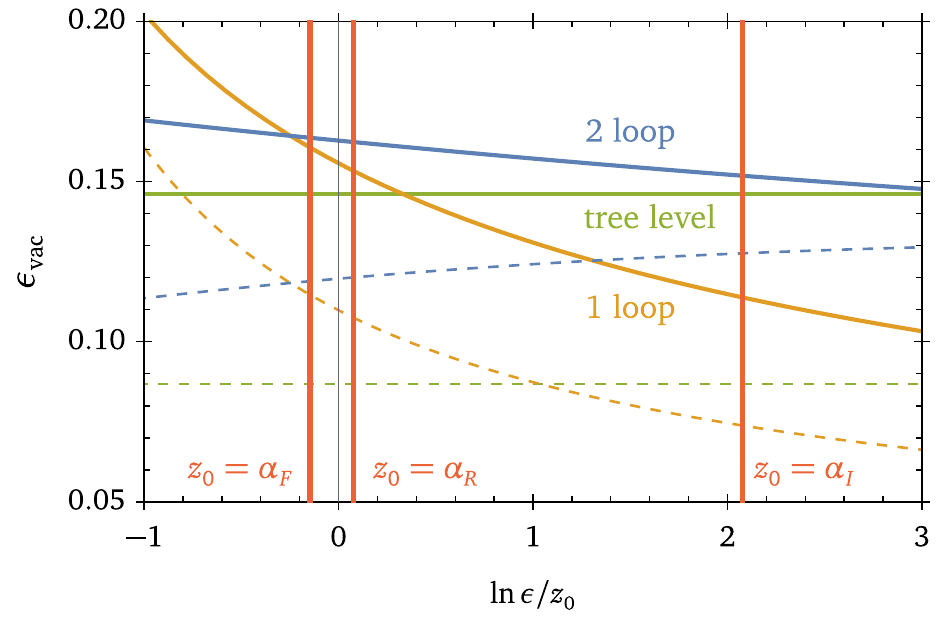}
    \caption{Upper bound of the conformal window $\epsilon_\text{vac}$ due to vacuum instability depending on the parameter $z_0$. The bound is extracted from the conformal expansion up to $\sim \epsilon^3$ (solid) and its $[2/1]$ Pad\'e resummation (dashed) of the effective potential at tree (green), one-loop (yellow) and two-loop (blue) level. Red lines mark special choices of $z_0$.}
    \label{fig:eps-vac-z0}
\end{figure}

The dependence of the upper values of the conformal windows on the parameter $z_0$ for tree-level, one and two loops in the effective potential is depicted in \fig{eps-vac-z0} (solid lines). We find both one- and two-loop corrections from the effective potential to predict a widening of the conformal window over the classical bound.
However, the two-loop result shows better convergence as it is less dependent on $z_0$, especially in the range \eq{z0-uncertainty} (vertical red lines).
We expect this trend to continue to higher loops where the $z_0$ dependence should diminish. 

\begin{figure}[h]
    \centering
    \includegraphics[width=0.66\textwidth]{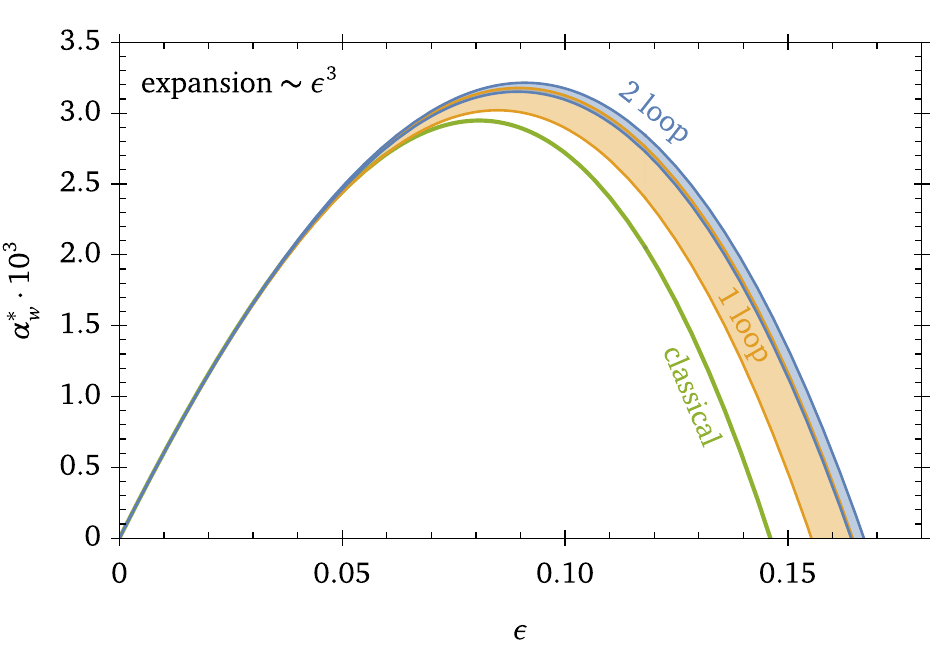}
    \caption{Effective fixed-point coupling $\alpha_w^*$ expanded up to $\propto \epsilon^3$, using only the classical (green), one- and two-loop approximation in the uncertainty range~\eq{z0-uncertainty} (yellow and blue bands). }
    \label{fig:aw-eps-1}
\end{figure}

The improvement of the fixed-point potential by quantum corrections in the $\epsilon$ expansion is depicted in \fig{aw-eps-1}.
Classical and improved potential differ notably starting around $\epsilon\approx 0.05$. 
Consequentially, the classical stability estimate $\epsilon^\text{exp}_\text{vac,cl} \approx 0.146$ is lifted decisively to
\begin{equation}\label{eq:res-exp}
    \epsilon^\text{exp}_\text{vac} \approx 0.166(1)
\end{equation}
at two loops. 
Both results are more stringent than the estimate of the single-trace merger $\epsilon_\text{mer,1} \approx 0.249$~\cite{Litim:2023tym}.  
The width of the uncertainty band of the effective potential is decreases from one to two loops. However, their overlap diminishes for larger $\epsilon$, slightly hinting at an inadequacy of the $\epsilon$ expansion at such values.

Finally, we note that an imaginary part of $\alpha_w$ is not manifest in the $\epsilon$ expansion. It would be heralded by $\Tilde{\alpha}_I < 0$~\eq{soft-aI}. However, we find that our resummation prescription 
\begin{equation}\label{eq:a_I-eps}
    \begin{aligned}
    \tfrac12 \Tilde{\alpha}_I  \approx   &\ 0.0625\,\epsilon \\
    &+ \left(0.0166 - 0.2438\,\ln \frac{\epsilon}{z_0}\right)\epsilon^2 \\
    &+ \left(0.0519 - 0.3509 \,\ln \frac{\epsilon}{z_0} - 0.0513\,\ln^2 \frac{\epsilon}{z_0}\right)\epsilon^3 + \mathcal{O}(\epsilon^4) \,,
    \end{aligned}
\end{equation}
is manifestly positive in the conformal expansion within the $z_0$-range~\eq{z0-uncertainty}.

\subsection{Pad\'e Resummations}

While the strict $\epsilon$ expansion is reliable for small $\epsilon$ and consistent with respect to higher-loop corrections, its convergence is rather poor. 
To remedy this, we employ Pad\'e resummations in $\epsilon$, serving as an estimate for the impact of higher powers in the $\epsilon$ expansion.
There are two approaches to this strategy. For one, the resummation may be applied to the fixed-point couplings $\alpha_{i=g,y,u,v}^* = \sum_{n=1}^3 c^{(i)}_n\,\epsilon^n$.
The relevant Pad\'e approximations read 
\begin{equation}
    [2/1]:\quad \alpha^* \approx c_1 \, \epsilon + \frac{c_2^2\,\epsilon^2}{c_2 - c_3\,\epsilon}\,, \qquad [1/2]:\quad \alpha^* \approx \frac{c_1^3 \,\epsilon}{c_1^2 + c_2^2\,\epsilon^2 - c_1  \epsilon (c_2 + c_3\,\epsilon)}\,.
\end{equation}
The second ansatz is to apply the $[2/1]$ Pad\'e resummation directly to $\alpha_w^*(\epsilon)$.
The $[1/2]$ approximant is not helpful as it remains stable by construction.
Note that both approaches are identical if only the classical potential is considered.

\begin{figure}[h]
    \centering
    \includegraphics[width=0.66\textwidth]{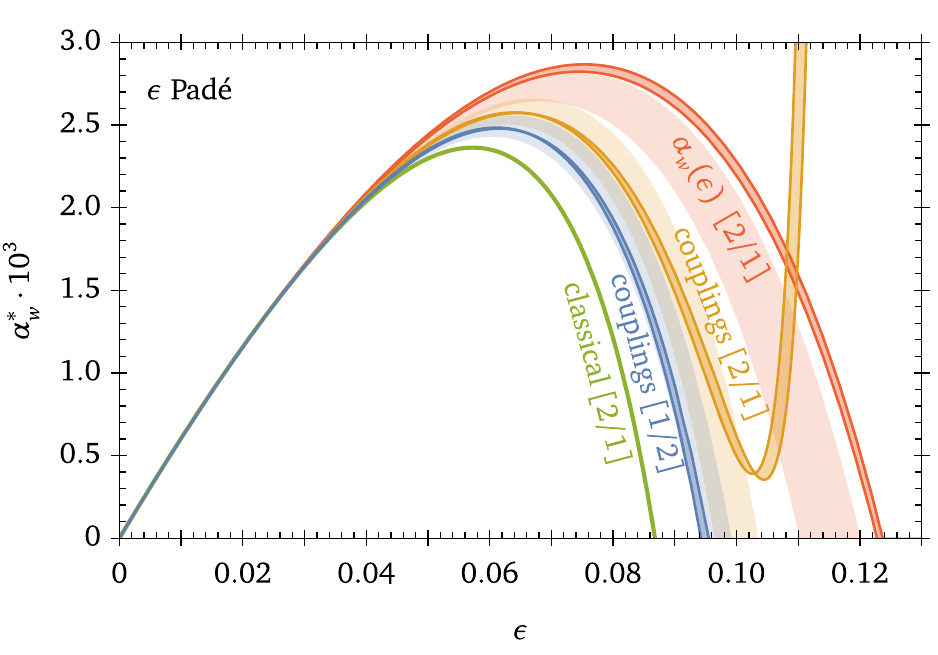}
    \caption{Pad\'e resummed fixed-point potential $\alpha_w^*$ in various approximations. This includes the $[2/1]$ approximants of the conformal expansion for the classical (green) and quantum effective potential (red), as well as the $[2/1]$ (yellow) and $[1/2]$ (blue) resummations of couplings $\alpha_{g,y,u,v}^*(\epsilon)$ as they contribute to $\alpha_w^*$. The $z_0$ uncertainty range \eq{z0-uncertainty} for one- and two-loop effective potentials are displayed as shaded regions.}
    \label{fig:aw-eps-2}
\end{figure}

The results are summarised in \fig{aw-eps-2}. As a reference, the $[2/1]$ of the classical potential with $\epsilon_\text{vac,cl}^{[2/1]} \approx 0.0868$ is shown in green. The corresponding resummation of the effective potential $\alpha_w^*$ is shown in red. 
Unfortunately, the prediction for the loss of vacuum stability appears to be under poor perturbative control since the two-loop prediction $0.1233(5)$ (darker shades) does not overlap with the one-loop result $0.115(5)$ (lighter shades) and should be disregarded.
Note that the $[2/1]$ approximation is depicted by dashed lines in \fig{eps-vac-z0}, highlighting that all resummations appear to predict a tighter conformal window than the $\epsilon$ expansion.  

Returning to \fig{aw-eps-2}, perturbative reliablility is slightly better when the Pad\'e resummation is applied to the couplings $\alpha_{g,y,u,v}^*$, where two-loop effective potentials are approximately contained within the one-loop uncertainties for most values of $\epsilon$. Using a $[2/1]$ resummation (yellow), the two-loop effective potential falls out of perturbative control around $\epsilon \approx 0.1$ and never enters an unstable regime. On the other hand, the $[1/2]$ approximant (blue) yields a two-loop prediction
\begin{equation}\label{eq:res-Pade}
    \epsilon_\text{vac}^\text{Pad\'e} \approx 0.0949(6)
\end{equation}
that is completely within the one-loop uncertainty. Note that this is larger than the single-trace merger $\epsilon_\text{mer,1}^{[2/1]} \approx 0.091$ obtained by the $[3/1]$ resummation of $\vartheta_1$~\cite{Litim:2023tym}.

\subsection{Resummation via \texorpdfstring{$\beta$}{Beta} Functions}

Another method to estimate higher powers in the $\epsilon$ expansion is to utilise the fixed-point conditions $\beta_{g,y,u,v} = 0$ to fix the couplings $\alpha_{g,y,u,v}^*$ exactly. 
As this approach is independent of the $\epsilon$ expansion, it is more sensitive to phenomena beyond strict perturbative control.
In this section, we will investigate these conditions at orders \texttt{lmn}, meaning \texttt{l} loops in the gauge, \texttt{m} in the Yukawa as well as \texttt{n} loops in the quartic $\beta$ functions. 
In order to expand fixed-point couplings up to $\propto \epsilon^\texttt{n-1}$ requires at least the loop combinations \texttt{(n+1)nn} and needs to be paired with the \texttt{(n-1)}-loop effective potential.

\begin{figure}[h]
    \centering
    \includegraphics[width=0.66\textwidth]{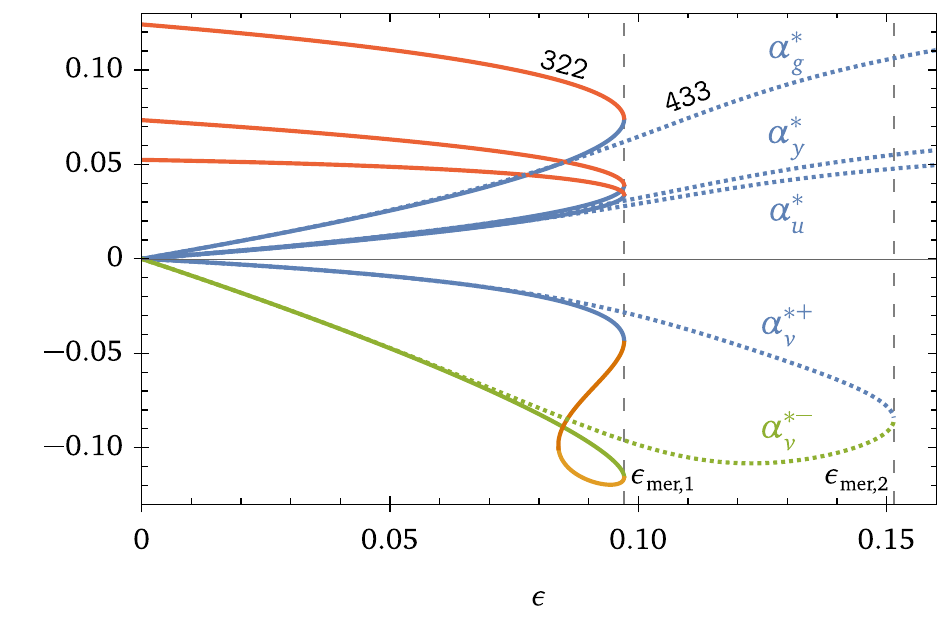}
    \caption{Fixed-point couplings $\alpha_{g,y,u,v}^*(\epsilon)$ in the \texttt{322} (solid) and \texttt{433} (dashed) approximation, showing the UV fixed point (blue), its second double-trace quartic (green), and other solutions (red, yellow, brown). The single-trace ($\epsilon_\text{mer,1}$) and double-trace merger ($\epsilon_\text{mer,2}$) of \texttt{322} and \texttt{433}, respectively, are marked by dashed vertical lines. }
    \label{fig:433vs322}
\end{figure}

Before discussing scalar stability, we recollect the findings of~\cite{Bond:2017tbw,Bednyakov:2023asy} regarding fixed-point mergers. 
\fig{433vs322} displays all fixed-point couplings $\alpha_{g,y,u,v}^*(\epsilon)$ both for \texttt{433} (dashed lines) and the previous order in the conformal expansion \texttt{322} (solid lines).
The UV fixed point is drawn in blue and its double-trace counterpart in green; both show good agreement between \texttt{433}  and \texttt{322} for small $\epsilon$. At \texttt{322}, however, the UV fixed point collides with an IR fixed point outside of strict perturbative control (red) via a single-trace merger at $\epsilon_\text{mer,1} \approx 0.0972$. On the other hand,  the conformal window at \texttt{433} persists up to a double-trace merger at $\epsilon_\text{mer,2} \approx 0.152$.

\begin{figure}[h]
    \centering
    \includegraphics[width=0.66\textwidth]{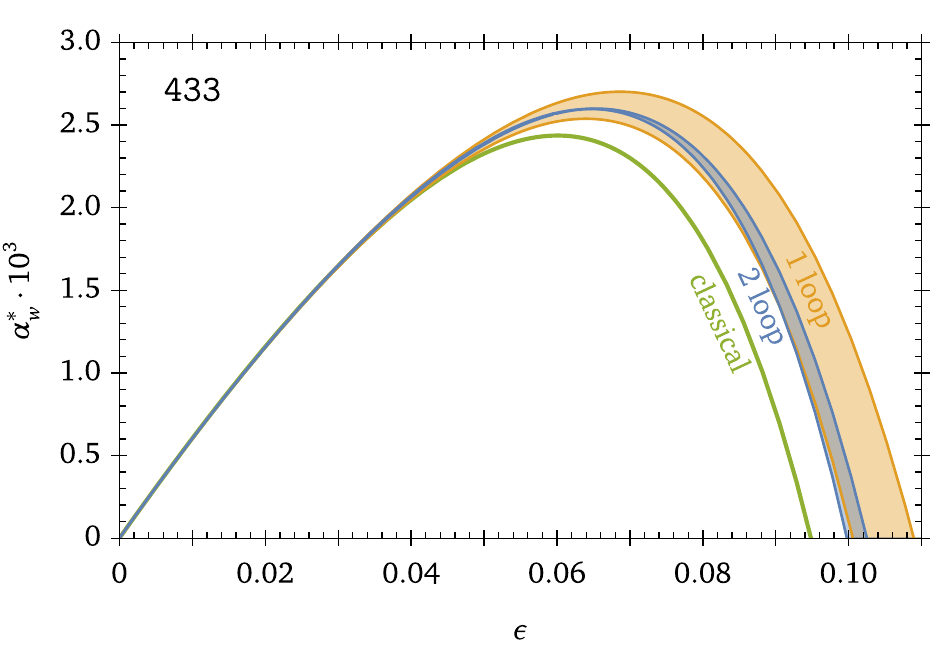}
    \caption{Fixed-point potential in the \texttt{433} approximation at tree- (green), one-loop (yellow) and two-loops (blue) level, with the $z_0$ uncertainty range \eq{z0-uncertainty} (shaded).}
    \label{fig:aw-eps-3}
\end{figure}

We now turn towards the stability analysis, where we will find tighter constraints than from the mergers. The tree-level and effective potential in the \texttt{433} approximation are displayed in \fig{aw-eps-3}. The two-loop uncertainty is much smaller and approximately contained within the one-loop one. It reads 
\begin{equation}\label{eq:res-433}
    \epsilon_\text{vac}^\texttt{433} \approx 0.1011(13)\,,
\end{equation}
which is a significant improvement over the tree-level result quoted in \cite{Bednyakov:2023asy}. In fact, the tree-level result is not even contained within the uncertainty range at one-loop.

\begin{table}[h]
    \centering
    \rowcolors{2}{LightGray}{}
\begin{tabular}{|c|lll|l|}
    \hline \rowcolor{LightBlue}
    Approx. & Tree level & 1 loop &  2 loop & Merger  \\
    \hline
    \texttt{433} & $0.0948$ & $0.1048(42)$ &  $0.1011(13)$ & $\epsilon_\text{mer,2} \approx 0.1515$ \\
    \texttt{432} & $0.0883$ & $0.0948(25)$ & $0.0929(5)$ &  $\epsilon_\text{mer,2} \approx 0.1364$ \\
    \texttt{423} & $0.1284$ & $0.169(23)$ & $0.1451(58)$ &  $\epsilon_\text{mer,1} \approx 0.4223$  \\
    \texttt{322} & $0.0867$ & $0.0913(16)$ & $0.090(4)$ & $\epsilon_\text{mer,1} \approx 0.0972$\\
    \hline
\end{tabular}
\caption{Bound on the UV conformal window at loop order \texttt{lmn}, from vacuum instability $(\epsilon_\text{vac})$, using the tree-level, one- and two loop effective potential as well as the merger. The uncertainty is estimated using \eq{z0-uncertainty}. }
\label{tab:lmn}
\end{table}

Furthermore, a range of relevant loop combinations for $\beta_{g,y,u,v}$ are collected in \tab{lmn}. In each case, the merger is only subdominant for the end of the conformal window.
Overall, the same picture as at \texttt{433} emerges: each approximation 
finds that the one- and two-loop potentials are within a perturbative agreement and admit roughly a $10\%$ widening of the conformal window with respect to the classical potential. In this sense, the resummation using $\beta$ functions appears to be more consistent than Pad\'e approximants.

\begin{figure}[h]
    \centering
    \includegraphics[width=0.66\textwidth]{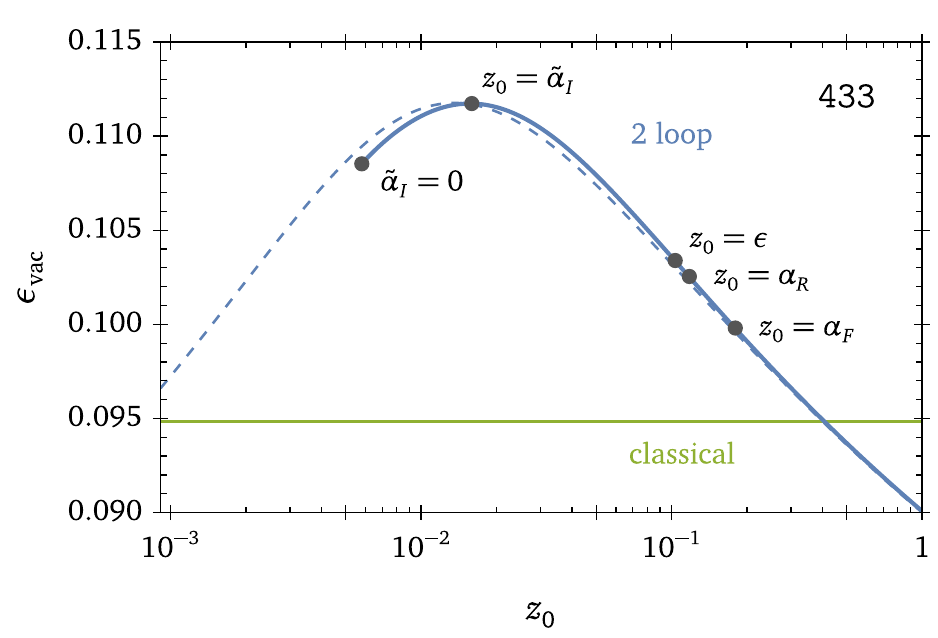}
    \caption{Upper bound for the UV conformal window $\epsilon_\text{vac}$ at \texttt{433} and its  $z_0$ dependence. The green line utilises only the classical stability condition. The solid blue curve takes into account the two-loop effective potential, and two-loop softening of $\tilde{\alpha}_I$ \eq{soft-aI}. The curve is only drawn for values of $z_0$ where $\text{Im}\,\alpha_w(\epsilon_\text{vac}) \neq 0$ as $\text{Re}\,\alpha_w(\epsilon_\text{vac}) = 0$.
    The dashed blue line does not employ the soft resummation~\eq{soft-aI}. It shows the bound from $\text{Re}\,\alpha_w(\epsilon_\text{vac})$ ignoring the imaginary part of $\alpha_w$ entirely.
    Several points highlight unique choices of $z_0$, the left and rightmost represent the ones taken into account for the uncertainty estimate \eq{z0-uncertainty}. Within this range, the prediction lies well above the classical one.}
    \label{fig:433-z0}
\end{figure}

Next we scrutinise the precision and reliability of our \texttt{433} prediction.
To that end, we study the  $z_0$-dependence of the stability prediction in more detail in \fig{433-z0} (blue curve) and contrast with the classical bound (green).
The impact of the soft resummation in $\tilde{\alpha}_I$ is displayed by comparing the resummed potential (solid) with the real part of the unresummed one (dashed), ignoring its imaginary part.
We observe that both agree well within the range \eq{z0-uncertainty}. We therefore expect that a different resummation prescription in the definition of $\tilde{\alpha}_I$ would have a similarly small impact.
Conversely, if $z_0$ inches closer towards $\tilde{\alpha}_I$, the difference between the resummed and unresummed potential increases. Also, the prediction for $\epsilon_\text{vac}$ increases significantly, showcasing the numerical inaccuracies for $z_0$ outside the range \eq{z0-uncertainty} as argued before.
Naturally, at $ z_0 \approx \tilde{\alpha}_I$ the prediction $\epsilon_\text{vac}$ is very sensitive to the definition of $\tilde{\alpha}_I$.
As $z_0$ is decreased even further, we eventually see the resummation prescription fail as the effective potential gains an imaginary part through $\tilde{\alpha}_I < 0$ while the real part stays positive.

\section{Discussion}\label{sec:Discussion}

In this work, the prediction for the scalar stability of the UV fixed point in the Litim-Sannino model has been significantly enhanced using loop corrections, renormalisation-group improvement, resummation of soft corrections and parameter optimisations.
In particular, the conformal window has been estimated using the conformal expansion \eq{res-exp} as well as resummations based on Pad\'e approximants \eq{res-Pade} and $\beta$ functions \eq{res-433}. In each of these approximations, we find a significant widening of the conformal window with respect to previous works~\cite{Litim:2023tym,Bednyakov:2023asy}.
As a result, smaller integer values of $(N_c,\,N_f)$ may be sufficient for a UV fixed point.

\begin{figure}[h]
    \centering
    \includegraphics[width=0.66\textwidth]{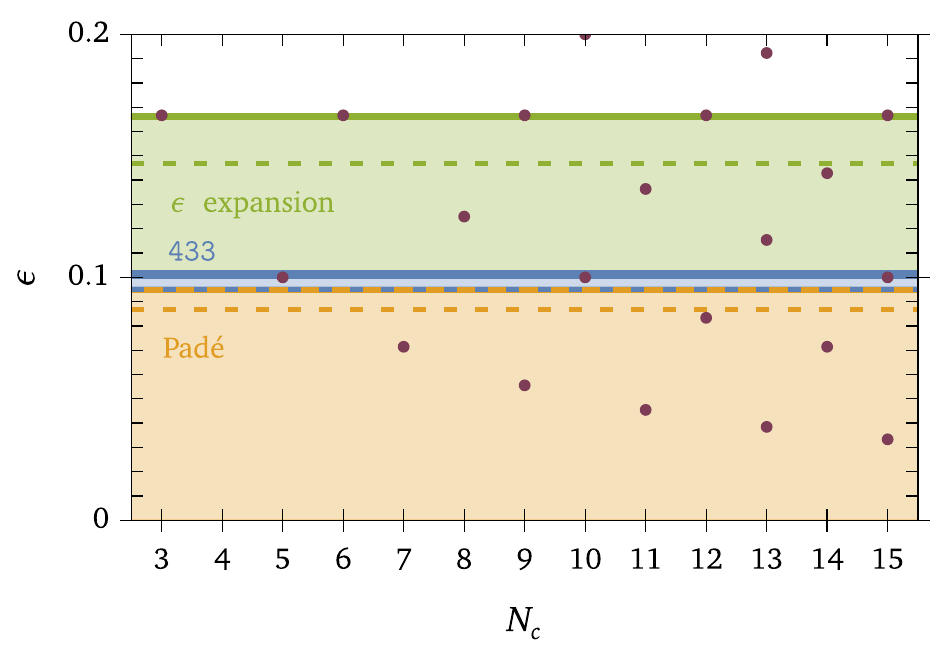}
    \caption{Conformal-window estimates from vacuum-instability estimates using the classical (dashed) or two-loop quantum improved potential (solid) in the conformal expansion (green), Pad\'e resummation (yellow) and \texttt{433} approximation. Dots mark integer choices of $N_{f,c}$ closest to the conformal window.}
    \label{fig:windows}
\end{figure}

This is displayed in \fig{windows}. Each dot represents an integer value of $N_{f,c}$, neglecting finite-$N$ corrections. The extend of the conformal window for each approximation is marked in different colours. The difference between dashed and solid lines mark the improvement due to quantum corrections of the effective potential.
It is only due to this enhancement that the conformal expansion points towards $(N_c,\,N_f) = (3,\,17)$ as the smallest multiplicity within the window. Similarly, the \texttt{433} approximation admits $(5,\,28)$ into the window. Neither of these theories were considered viable in \cite{Litim:2023tym,Bednyakov:2023asy}.

In our analysis, we have verified that uncertainties in the stability predictions overall decrease with higher loops in the effective potential. Moreover, we have used the consecutive nestedness of these regions as sanity criterion for our approximations. 

However, it is not possible to decide conclusively which bound in \fig{windows} is most compelling with the information at hand.
The root cause is the remaining uncertainty of fixed-point couplings due to unknown higher-loop corrections of their $\beta$ functions.
While the fixed point is most universally described by its conformal expansion, the power series~\eq{a_lambda-eps} is very short. In fact, we find soft hints that the convergence of the conformal expansion is too weak for \eq{res-exp} to be a reliable prediction of vacuum instability at the current order.
Instead, resummations in $\epsilon$ appear to be more credible. In particular, both \eq{res-Pade} and \eq{res-433} seem to be in agreement that the conformal window is substantially smaller than \eq{res-exp}.
Each prediction represents an educated guess for higher powers in $\epsilon$, with many alternative resummation strategies being conceivable. As it stands, it cannot be assessed which approach is most accurate.
We expect to find increasingly better agreement between conformal expansion and resummation techniques with higher loop orders, which will eventually allow a more conclusive prediction of the UV conformal window.

Moreover, the possible occurrence of a fixed-point merger cannot be ruled out. This phenomenon may lead to the breakdown of UV conformal window even before vacuum stability is lost.
Mergers are notoriously difficult as they cannot be captured within the conformal expansion of fixed point couplings, where they cause a sharp decline of convergence. Instead, they show up in critical exponents which need to be determined with high confidence.
Furthermore, the colliding IR fixed point might not be under strict perturbative control as $\epsilon \to 0$. Both properties can be seen in \fig{433vs322} (red curve). 

In summary, while this work suggests an enhancement of the UV conformal window for each of the approximations in \fig{windows}, a confident conclusion about its size and breakdown mechanism cannot be reached. 
Higher loops or non-perturbative arguments are required to settle these questions.

\section*{Acknowledgement}
The author is indebted to Emmanuel Stamou for valuable discussions and comments on the manuscript.

\bibliographystyle{JHEP}
\bibliography{ref.bib}
\end{document}